\documentclass[letterpaper, 10 pt, conference]{ieeeconf}  

\IEEEoverridecommandlockouts                              
\usepackage{cite}

\usepackage{amsmath,amssymb,amsfonts,amsthm}
\usepackage{graphicx}
\usepackage{textcomp}
\usepackage{tabularx}
\usepackage[capitalize]{cleveref} 
\usepackage{multirow}

\usepackage{mathtools}

\usepackage{optidef}

\usepackage{float}
\usepackage{dblfloatfix}
\usepackage{caption}
\usepackage{subcaption}
\usepackage[dvipsnames]{xcolor}
\restylefloat{table}

\usepackage[ruled,vlined]{algorithm2e}
\usepackage{algorithmic}

\usepackage{pgfplots}
\DeclareUnicodeCharacter{2212}{−}
\usepgfplotslibrary{groupplots,dateplot}
\usetikzlibrary{patterns,shapes.arrows}
\pgfplotsset{compat=newest}

\DeclareMathAlphabet{\mymathbb}{U}{BOONDOX-ds}{m}{n}

\newcommand{\beeq}{\begin{equation}}
\newcommand{\eneq}{\end{equation}}
\newcommand{\matb}{\begin{matrix}}
\newcommand{\mate}{\end{matrix}}

\DeclareMathOperator*{\Tr}{Tr}
\DeclareMathOperator*{\diag}{diag}

\newcommand{\Epsilon}{\mathcal{E}}

\usepackage{tabularx} 
\usepackage{booktabs} 

\usepackage{mathrsfs}

\usepackage{cuted}

\let\labelindent\relax 
\usepackage[shortlabels]{enumitem}

\newtheorem{theorem}{Theorem}

\newtheorem{lemma}{Lemma}
\newtheorem{assumption}{Assumption}

\theoremstyle{definition}
\newtheorem{definition}{Definition}
\theoremstyle{plain}
\newtheorem{remark}{Remark}

\title{\LARGE \bf
The Bias of Subspace-based Data-Driven Predictive Control}

\author{Keith Moffat, Florian Dörfler, and Alessandro Chiuso
\thanks{\noindent This research is supported by the Swiss National Science Foundation under the NCCR Automation (grant agreement 51NF40\_225155) and by the Italian Ministery of Reseaerch (PRIN “Data-driven learning of constrained control systems”, contract no. 2017J89ARP).}
\thanks{K. Moffat and F. Dörfler are with the Automatic Control Laboratory, Swiss Federal Institute of Technology (ETH) Z\"urich, Switzerland. (e-mail: {\tt\small \{kmoffat, dorfler\}@ethz.ch})}%
\thanks{Alessandro Chiuso is with the Dept. of Information Engineering, University of Padova, Italy. (e-mail: {\tt\small alessandro.chiuso@unipd.it})}%
}

\begin{document}

\maketitle
\thispagestyle{empty}
\pagestyle{empty}





\begin{abstract}
    This paper quantifies and addresses the bias of subspace-based Data-Driven Predictive Control (DDPC) for linear, time-invariant (LTI) systems. 
    The primary focus is the bias that arises when the training data is gathered with a feedback controller in closed-loop with the system.
    First, the closed-loop bias of Subspace Predictive Control is quantified using the training data innovations. 
    Next, the bias of direct, subspace-based DDPC methods DeePC and $\gamma$-DDPC is shown to consist of two parts---the Subspace Bias, which arises from closed-loop data, and an Optimism Bias, which arises from DeePC/$\gamma$-DDPC's ``optimistic'' adjustment of the output trajectory.
    We show that, unlike subspace-based DDPC methods, Transient Predictive Control does not suffer from Subspace Bias or Optimism Bias.
    Double integrator experiments demonstrate that Subspace and Optimism Bias are responsible for poor reference tracking by the subspace-based DDPC methods.
\end{abstract}


\section{Introduction}\label{sec:introduction}

Direct Data-Driven Predictive Control (DDPC) has recently demonstrated impressive performance \cite{coulson2019data, berberich2020data}, with the literature focusing on the importance of regularization when the data is noisy \cite{coulson2021distributionally, huang2023robust, breschi2022unified, chiuso2025harnessing}.
This paper focuses on an additional important aspect of DDPC---the bias of subspace-based DDPC for Linear, Time-Invariant (LTI) systems.
We distinguish two types of bias for subspace-based DDPC---``Subspace Bias,'' which arises from closed-loop data, and ``Optimism Bias,'' which arises from the ``optimistic'' adjustment of the subspace-predicted output trajectory.

While Optimism Bias is a new term, Subspace Bias has been described in the literature.
Often, in practice, it is necessary to gather data for DDPC using an existing, suboptimal feedback controller. 
However, issues can arise when the training-time controller is replaced by the DDPC.
The challenge of closed-loop data for Subspace Predictive Control (SPC) \cite{FAVOREEL1999} was pointed out in \cite{dong2008CLSPC, huang2008dynamic}, and recently in reference to Data-enabled Predictive Control (DeePC) \cite{coulson2019data} in \cite{dinkla2023bias}, which derived an expression for the bias of the Subspace Predictor parameters.

Prior to the attention from the DDPC community, the system identification community recognized that closed-loop data biases subspace system identification of LTI systems \cite{Vanov-book, ChiusoAUTO06, vanderveen2013closed}. 
The system identification bias has been addressed using single-step predictors \cite{LjungMcK1996, OverscheeDeMoor97, ChiusoP-05a}, Instrumental Variables (IVs) \cite{pouliquen2010indirect},
and innovation estimation \cite{QinL2003a, ChiusoP-05a}.

Theorem \ref{thm:bias}, presented in this paper, quantifies the bias of subspace-based mulstistep predictions when the training data is gathered in closed loop, while Theorem \ref{thm:biasDDPC} quantifies the closely-related biases of DeePC \cite{coulson2019data} and $\gamma$-DDPC \cite{breschi2022unified}.
The bias expressions in Theorems \ref{thm:bias} and \ref{thm:biasDDPC} include the training data innovations, which can be estimated a posteriori (after the data are gathered but before the controller is used). 

In contrast to subspace-based DDPC methods (SPC, DeePC, and $\gamma$-DDPC), 
and following the insight in \cite{LjungMcK1996} that single-step predictors are not biased by closed-loop data, a consistent multistep predictor can be assembled by sequentially applying single-step predictors into the future.
Furthermore, a single-step predictor bank can be reformulated to produce a consistent estimate of the ``true'' certainty-equivalent Multistep Predictor, as demonstrated in \cite{moffat2024transient}.

Two single-step predictor methods exist in the literature---the ``Fixed-Length'' and ``Transient'' Predictors \cite{moffat2024transient}. 
The Fixed-Length Predictor constructs the single-step predictor bank from a single ARX model repeated and staggered in time \cite{dong2008CLSPC, dinkla2024closed} and is equivalent to the certainty-equivalent portion of the method in \cite{chiuso2025harnessing} which, in addition, derives a separation principle for DDPC and an optimal input regularization.
While this paper uses the Transient Predictor for comparison with the subspace-based methods, the Fixed-Length Predictor performs comparably-well on the experiments in this paper. 
For a comparison between the Transient and Fixed-Length Predictors, see \cite{moffat2024transient}.

\noindent The contributions of this paper are, for LTI systems,
\begin{enumerate}
    \item The definitions of Subspace Bias and Optimism Bias,
    \item {\setlength{\spaceskip}{3pt}Theorems \ref{thm:bias} and \ref{thm:biasDDPC},} which quantify the bias of the Subspace Predictor and DeePC/$\gamma$-DDPC, respectively, and 
    \item experiments demonstrating the impact of bias on DDPC and the consistency of the Transient Predictor.
\end{enumerate}


\section{Preliminaries}\label{sec:preliminaries}

\vspace{-5pt}

\subsection{Notation}\label{sec:notation}
 
Given a matrix A, its transpose is $A^T$, 
its Frobenius norm is $\left\Vert A \right\Vert_F$,
its Cholesky factorization is $A^{1/2}$, and the Cholesky factorization of its inverse $A^{-1}$ is $A^{-1/2}$.
$\sigma_{\text{max}}(A)$ is the largest singular value of $A$.
The $d$-dimensional identity matrix is $I_d$. 
The 2-norm is denoted by $\left\Vert \cdot \right\Vert$. 

The number of columns of a Hankel matrix is $N$.
For any signal $w(t) \in \mathbb{R}^{n_w}$, we define the associated $\frac{1}{\sqrt{N}}$-scaled Hankel matrix
$W_{[t_0,t_1]} \in \mathbb{R}^{n_w(t_1-t_0+1) \times N}$ as:
\smaller
\begin{align*}
    W_{[t_0,t_1]} := \frac{1}{\sqrt{N}}\begin{bmatrix}
        w(t_0) & w(t_0 + 1) & \hdots & w(t_0 + N - 1) \\
        w(t_0+1) & w(t_0 + 2) & \hdots & w(t_0 + N ) \\
        \vdots & \vdots & \ddots & \vdots \\
        w(t_1) & w(t_1 + 1) & \hdots & w(t_1 + N - 1) \\
    \end{bmatrix},
\end{align*}
\normalsize
where the $\frac{1}{\sqrt{N}}$ scaling normalizes the variance.

\subsection{System Class}\label{sec:systemClass} 

We consider the discrete-time, LTI system class, whose input $u(t) \in \mathbb{R}^m$ and output $y(t) \in \mathbb{R}^q$ constitute a stationary joint process
\begin{align*}
    z(t) = \begin{bmatrix} y(t) \\ u(t)  \end{bmatrix} \in \mathbb{R}^{q+m}.
\end{align*}
The stationary joint process can be represented using an infinite-length/steady-state ARX representation:
\begin{align}\label{eqn:sysDef}
    y(t+1) &= \sum_{i=1}^{\infty} \phi_{\infty, i} z(t-i+1) + e(t+1),
\end{align}
with $\phi_{\infty, i} \in \mathbb{R}^{q \times (q+m)}$. $e$ is zero-mean white noise which results in prediction errors, or the \emph{stationary innovation}.

Given a finite past-horizon of length $l$, we have a finite-length/transient ARX representation:
\begin{align}\label{eqn:sysDefFinite}
    y(t+1) = \sum_{i=1}^{l} \phi_{l,i} z(t-i+1) + \epsilon_{l}(t+1),
\end{align} 
where $\phi_{l,i} \in \mathbb{R}^{q \times (q+m)}$ is the $i^\text{th}$ block-entry of the $l$-long optimal (minimum variance, linear, unbiased) ``transient'' single-step predictor.
$\epsilon_{l}$ is the \emph{transient innovation} for $l$ \emph{lead-in} data \cite{moffat2024transient}.

\subsection{Variables and definitions} 

The future, $\tau$-long input and output sequences are 
\begin{align*}
    u_f := u_f(t) &:= \begin{bmatrix} u^T(t+1) & \hdots & u^T(t + \tau) \end{bmatrix}^T \text{ and} \\
    y_f := y_f(t) &:= \begin{bmatrix} y^T(t+1) & \hdots & y^T(t + \tau) \end{bmatrix}^T.
\end{align*}
The $\rho$-long lead-in measurement sequence is
\begin{align*}
    z_p := z_p(t) &:= \begin{bmatrix} z^T(t-\rho+1) & \hdots & z^T(t) \end{bmatrix}^T \in \mathbb{R}^{(q+m)\rho},
\end{align*}
which encodes the initial condition at time $t$. 
The innovation sequence for the future $\tau$ steps with $\rho$ lead-in data is
\begin{align*}
    \epsilon_f := \epsilon_f(t) &:= \begin{bmatrix} \epsilon_\rho^T(t+1) & \hdots \epsilon_{\rho+\tau-1}^T(t + \tau) \end{bmatrix}^T \in \mathbb{R}^{q\tau}.
\end{align*}

For notational convenience, we define the Hankel matrices 
$Z_p := Z_{[1,\rho]}$, 
$U_f :=  U_{[\rho+1,\rho+\tau]}$,
$Y_f :=  Y_{[\rho+1,\rho+\tau]}$, and
$\Epsilon_f :=  \Epsilon_{[\rho+1,\rho+\tau]}$.
We also define the Multisptep Predictor matrices $H_p \! \in \! \mathbb{R}^{q \tau \times (q+m)\rho } \! $, $H_u \! \in \! \mathbb{R}^{q \tau \times m\tau} \! $, and $H_\epsilon \! \in \! \mathbb{R}^{q \tau \times q \tau} \! $ such that
\begin{align*} 
    y_f &= H_p z_p + H_u u_f + H_\epsilon \epsilon_f.
\end{align*} 
Removing the innovation term, which cannot be anticipated, the certainty-equivalent multistep output prediction is
\begin{align}\label{eqn:optMultiPred}
    H v 
    := \begin{bmatrix}
        H_p & H_u
    \end{bmatrix} \begin{bmatrix}
        z_p \\ u_f
    \end{bmatrix}.  
\end{align}
$H$ is the Multistep Predictor for $\tau$ steps into the future and 
$$ v := \begin{bmatrix} z_p \\ u_f \end{bmatrix} \in \mathbb{R}^{(q+m)\rho + m\tau}.$$

We also define the Subspace Predictor 
\begin{align*}
    S := \mathbb{E}\left[Y_f V^T\right] \mathbb{E}\left[V V^T\right]^{-1}, \text{ where} 
\end{align*}
\begin{align*}
    V &:= \begin{bmatrix} Z_p \\ U_f \end{bmatrix} \in \mathbb{R}^{((q+m)\rho + m\tau) \times N},
\end{align*}
and its sample estimate $\widehat{S} := Y_f V^T \left( V V^T\right)^{-1}.$

\subsection{Data Collection}\label{sec:asssumptions}

We define the ``training-time feedback controller'' $f(\cdot)$ such that the closed-loop input 
\begin{align}\label{eqn:feedbackEqn}
    u(t) = f\big(z_{-\infty}(t)\big) + \alpha(t),
\end{align}
where $z_{-\infty}(t) := \begin{bmatrix} z^T(t-\infty) & \hdots & z^T(t) \end{bmatrix}^T$ and $\alpha(t)$ is the non-feedback portion of the input. We note that $f(z_{-\infty}(t))$ is general and allows for, e.g., infinite-past-horizon Kalman-filtered state feedback as well as ``immediate output feedback'' in which $u(t) = f\big(y(t)\big)$.

\begin{remark}\label{rem:contig}
$f(\cdot)$ is ``external'' to the system \eqref{eqn:sysDef} and thus its effect is not included in the Multistep Predictor $H$ \eqref{eqn:optMultiPred}.
\end{remark}

In addition, we make the following assumption on the training data input and the noise
to guarantee that the data can provide a consistent estimate of $H$. 


\begin{assumption}\label{ass:peristency}
The input $u(t)$ used to build the training data is persistently exciting such that, for some positive $c$,
\begin{align*}
    U_{[1,\rho+\tau-1]} U_{[1,\rho+\tau-1]}^T \geq cI_{\rho+\tau-1} \: \forall \: N, 
\end{align*}
and the variance of the noise $e(t)$ is positive definite.
\end{assumption}
Assumption \ref{ass:peristency} guarantees that that $Z_{[1,l]} Z_{[1,l]}^T$ does not become singular as $N \rightarrow \infty$ for any $l \in [\rho, \rho+\tau-1]$ and thus, given sufficient data, the single-step predictors can be estimated with arbitrary precision using \eqref{eqn:singleStepPred}.




\section{The Bias of Subspace Predictive Control}\label{sec:SPCbias}


SPC determines inputs by solving the optimization
\begin{mini}|s| 
{u_f,y_f}{J(u_f,y_f)}
{}{}
\addConstraint{y_f = \widehat{S}v}.
\label{eqn:SPCDef}
\end{mini}
This section derives the bias of $S$ when training data used to construct $Y_f$ and $V$ is gathered in closed loop, i.e., using \eqref{eqn:feedbackEqn}.
The bias arises from SPC ``misinterpreting'' the causality of the relationship between $\epsilon_f$ and $u_f$ and trying to use $u_f$ to set $\epsilon_f$, which is not possible.
 \begin{figure}
      \centering
      \begin{subfigure}[b]{.48\columnwidth}
          \centering
          \includegraphics[width=\columnwidth]{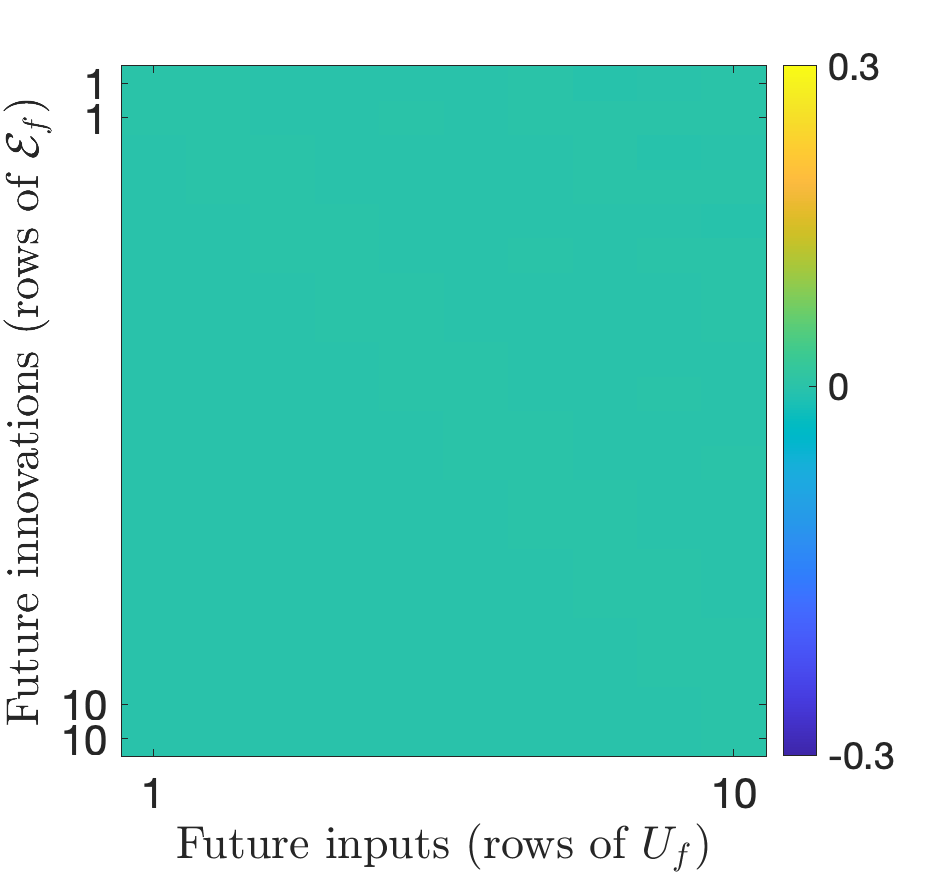}
          \caption{Without feedback}
          \label{fig:corrWoFeedback}
      \end{subfigure}
      \hfill
      \begin{subfigure}[b]{.48\columnwidth}
          \centering
          \includegraphics[width=\columnwidth]{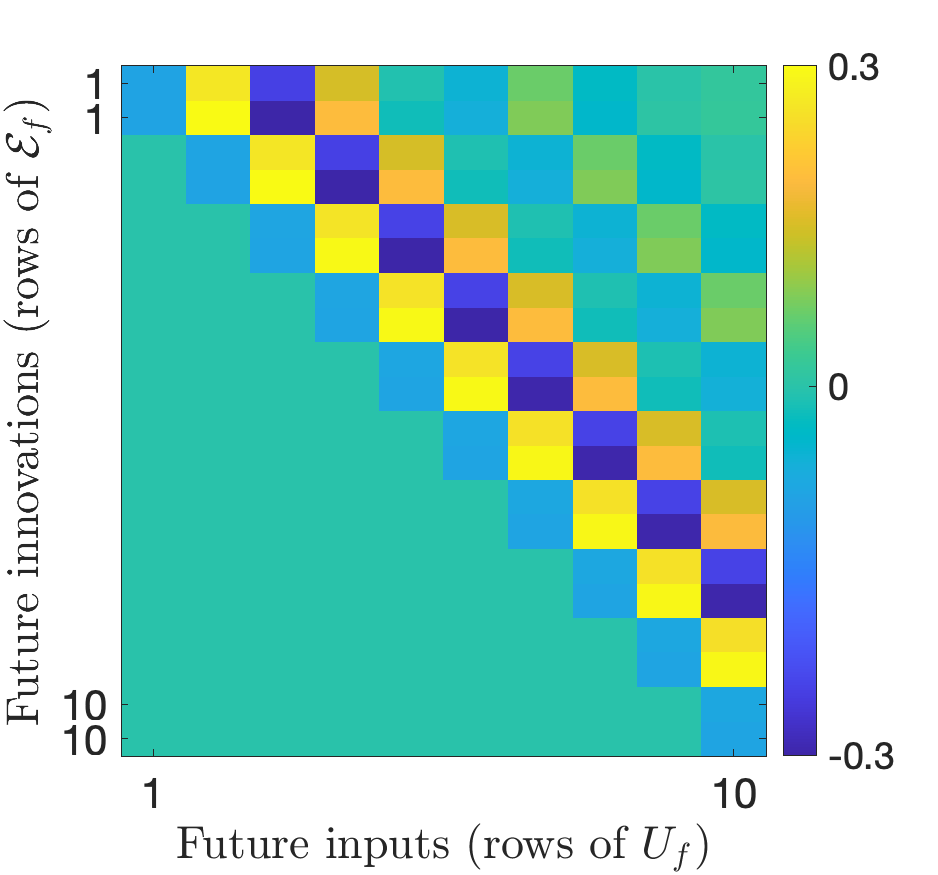}
          \caption{With feedback}
          \label{fig:corrWfeedback}
      \end{subfigure}
      
      \caption{\small Double integrator input-innovation correlation averaged over 100 independent experiments. \\ \smaller (The $y$-axis entries are duplicated because there are two outputs.)\normalsize}
      \label{fig:corr}
 \end{figure}

\begin{definition}
    The (LTI) \emph{Subspace Bias} for a given $v$ is
    \begin{align*}
        b_{S}(v) := Sv - Hv \in \mathbb{R}^{q \tau}.
    \end{align*}
\end{definition}

\begin{lemma}\label{lemma:SPCBias}
    The Subspace Bias is 
    \begin{align}\label{eqn:biasDef}
        b_{S}(v) = \beta v,
    \end{align}
    where we define Subspace Bias Predictor $\beta$ as
    \begin{align}\label{eqn:betaExpression}
        \beta := H_\epsilon \mathbb{E}\big[\Epsilon_f U_f^T\big] M \in \mathbb{R}^{((q \tau) \times (q+m)\rho + m \tau)}
    \end{align} 
    and the weighting matrix $M$ is given in \eqref{eqn:Mdef}. 
\end{lemma}
Lemma \ref{lemma:SPCBias}, proven in Appendix \ref{app:Proofs}, demonstrates that the Subspace Bias depends upon the correlation between the inputs and the innovations.
Fig. \ref{fig:corr} plots the correlation for 1000 data points from a double integrator, the Euler Discretization of $y = \ddot{x}$, 
with Gaussian measurement noise $\sim \mathcal{N}(0,0.0004)$.
The feedback law for Fig. \ref{fig:corrWfeedback} is the proportional-derivative feedback law $f\big(y(t)\big) = -2.5 y_1(t) - 3y_2(t)$, where $y_1(t)$ and $y_2(t)$ are the position and velocity of the mass, respectively, and $\alpha(t)$ is random Gaussian excitation $\sim \mathcal{N}(0,0.01)$. 
It is evident that $\Epsilon_f U_f^T \neq 0$ when the data is gathered in closed loop as 
$\epsilon(k)$ affects $u(j)$ for $j \geq k$.


\begin{remark}\label{rem:explainSPCbias}
    The SPC bias comes from the correlation between the innovation data $\Epsilon_f$ and the input data $U_f$. 
    SPC does not assert causality and interprets the correlation between $\Epsilon_f$ and $U_f$ in an anti-causal manner---the SPC prediction tries to use $u(t+j)$ to affect $\epsilon(t+k)$ for $j \geq k$. Of course, this does not work and, instead, introduces a bias.

    We note, too, that the closed loop bias of SPC only arises when the innovation is nonzero. For deterministic systems with $e(t) = 0$, there is no innovation for SPC to misinterpret, $\Epsilon_f = 0$, and $b_{S}(v) = 0$. 
\end{remark}

Before introducing Theorem \ref{thm:bias}, we introduce Subspace Empirical Bias, which clarifies the connection between the SPC Bias in this section with the $\gamma$-DDPC/DeePC bias in Section \ref{sec:DDPCbias}.
The (LTI) \emph{Subspace Empirical Bias} for the training data is
\begin{align*}
    \beta_V := SV - HV = \beta V \in \mathbb{R}^{q \tau \times N}.
\end{align*}

\begin{lemma}\label{lemma:spcBiasLinComb}
    The Subspace Bias for any $v$ is a linear combination of the Subspace Empirical Bias:
    $$b_{S}(v) = \beta_V g \text{ with }g \text{ s.t. } v = V g.$$
\end{lemma}

Theorem \ref{thm:bias} quantifies the maximum bias and the ``expected bias magnitude'' of the Subspace Predictor. 

\begin{theorem}\label{thm:bias} 
    $\\$ 
    Consider an LTI system, i.e., \eqref{eqn:sysDef}, 
    
    \noindent a) If training data is collected in open loop,
    $ b_{S}(v) = 0$.

    \noindent b) If the training data is gathered using feedback, 
    \begin{align*} 
        \begin{aligned} 
        &\underset{v,g}{\text{max}} \;\; \left\lVert b_{S}(v) \right\lVert \\
        &\text{s.t.} \;\ v = V g \text{, } \|g\| = 1  
        \end{aligned} 
        \quad = \sigma_\text{max} \left( \beta_V \right) \neq 0,
    \end{align*}
    where $\beta_V$ is the Subspace Empirical Bias.
    Furthermore, defining $p(V)$ as the distribution from which $V$ was drawn,
    \begin{align*}
        \!\mathbb{E}_{v \sim p(V)} \!\! \left[ \left\Vert b_{S}(v)\right\Vert^2\right] \!&=\! \left\Vert H_\epsilon \! \begin{bmatrix}
            0 & \mathbb{E}\big[\Epsilon_f U_f^T\big]
        \end{bmatrix} \mathbb{E}\left[ V V^T \right]^{-1/2}\right\Vert^2_F \!.
    \end{align*}
\end{theorem}

Theorem \ref{thm:bias} states that the Subspace Predictor is unbiased when the training data is collected in open loop, but not, in general, when the data is collected in closed loop.
Specifically, the first part of Theorem \ref{thm:bias}b) provides a worst-case bias magnitude 
which is important for optimization-based control because the optimizer may ``seek out'' the behavior described by the bias. 
The second part of Theorem \ref{thm:bias}b) gives the ``expected bias magnitude''---the Subspace Bias if $v$ is drawn randomly from the same distribution as the training data $V$. While $v$ will not be drawn from the same distribution as $V$ in DDPC applications, the expected bias magnitude provides a ballpark-range for the Subspace Bias. 



\begin{figure}[h]
     \centering
     \begin{subfigure}[b]{.48\columnwidth}
         \centering
         \includegraphics[width=\columnwidth]{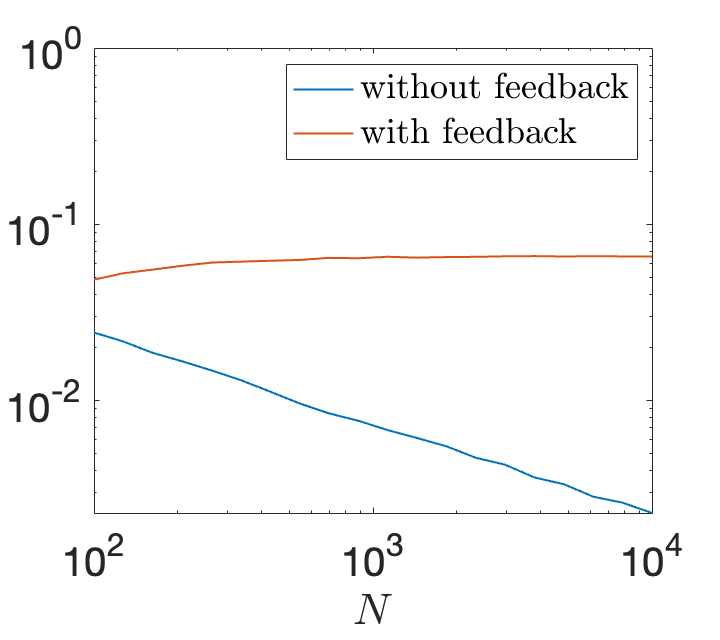}
         \caption{\smaller{$\sigma_\text{max} \left( \widehat{\beta_V} \right)$}}
         \label{fig:largestSingVal}
     \end{subfigure}
     \hfill
     \begin{subfigure}[b]{.48\columnwidth}
         \centering
         \includegraphics[width=\columnwidth]{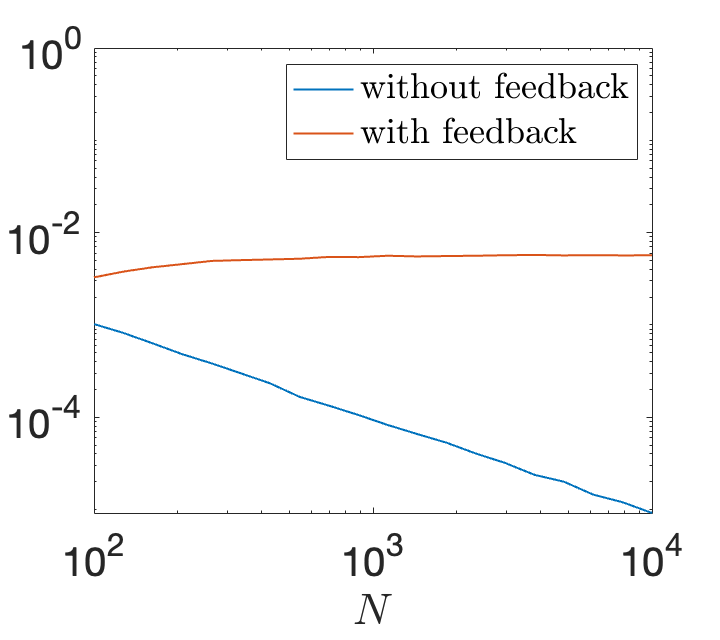}
         \caption{\smaller{$\mathbb{E}_{v \sim p(V)}\left[ \left\Vert \widehat{b}_{S}(v)\right\Vert^2\right]$}}
         \label{fig:frobNorm}
     \end{subfigure}
     \caption{\small Asymptotic characteristics of the bias-defining quantities in Theorem \ref{thm:bias}, averaged over 100 experiments.} 
     \label{fig:expectedBiasAsymptotics}
\end{figure}

Fig. \ref{fig:expectedBiasAsymptotics} plots the bias-defining quantities from Theorem \ref{thm:bias} as a function $N$ for the same double integrator experiment as Fig. \ref{fig:corr}. 
It demonstrates that when there is feedback in the data, 
both the largest singular value and the expected bias in Theorem \ref{thm:bias} converge to non-zero values. i.e., there will be prediction bias regardless of how much data is gathered.

In practice, $\beta$ is not available and must be estimated from data. 
Section \ref{sec:estimatingB} describes how to do this.

\section{The Bias of DeePC and $\gamma$-DDPC}\label{sec:DDPCbias}

The DeePC \cite{coulson2019data} optimization is 
\begin{align}
    \underset{u_f, y_f, g}{\min} \quad & J(u_f, y_f, g) \label{eqn:optiDef_DeePC} \\
    \text{s.t.} \quad \quad &\begin{bmatrix}
                    v \\ y_f
                \end{bmatrix} =\begin{bmatrix}
                    V \\ Y_f
                \end{bmatrix} g. \nonumber
\end{align}
The $\gamma$-DDPC \cite{breschi2022unified} optimization is 
\begin{align}
    \underset{u_f, y_f, \gamma_v, \gamma_{\epsilon}}{\min} \quad & J(u_f, y_f, \gamma_v, \gamma_{\epsilon}) \label{eqn:optiDef_gammaDDPC}\\
    \text{s.t.} \quad \quad &\begin{bmatrix}
                    v \\ y_f
                \end{bmatrix} =
                \begin{bmatrix}
                    L_{VV} & 0 \\
                    L_{Y V} & L_{Y \Epsilon}
                \end{bmatrix}
                \begin{bmatrix}
                    \gamma_v \\ \gamma_{\epsilon}
                \end{bmatrix}, \nonumber
\end{align}
which uses the LQ decomposition of the training data,
\begin{align}\label{eqn:LQdecompData}
    \begin{bmatrix}
        V \\ Y_f
    \end{bmatrix} &=: LQ =: 
    \begin{bmatrix}
        L_{VV} & 0 \\
        L_{Y V} & L_{Y \Epsilon}
    \end{bmatrix}
    \begin{bmatrix}
        Q_V \\ Q_{\Epsilon}
    \end{bmatrix}.
\end{align}



In this section we present an expression for the bias of the prediction implicit in $\gamma$-DDPC (DeePC) for a particular choice of $\gamma := \begin{bmatrix} \gamma_v^T & \gamma_{\epsilon}^T \end{bmatrix}^T$ ($g$).
First, we note that the Subspace Predictor is embedded in the LQ decomposition: 
\begin{align*}
    \mathbb{E}\left[L_{Y V}\right] \mathbb{E}\left[L_{VV}\right]^{-1} = \mathbb{E}\left[Y_f V^T\right] \mathbb{E}\left[V V^T\right]^{-1} = S.
\end{align*}

The (LTI) \emph{$\gamma$-DDPC Bias} for a $\gamma$ from \eqref{eqn:optiDef_gammaDDPC} is 
\begin{align*}
    b_{\gamma}(\gamma) := \begin{bmatrix}
        L_{Y V} & L_{Y \Epsilon}
    \end{bmatrix} \gamma 
    - H L_{VV} \gamma_v,
\end{align*}
and the (LTI) \emph{DeePC Bias} for a $g$ from \eqref{eqn:optiDef_DeePC} is
\begin{align*}
    b_{g}(g) := Y_f g 
    - H V g.
\end{align*}

\begin{lemma}\label{lemma:DDPCbias}
    The $\gamma$-DDPC Bias as a function of $\gamma$ is 
    \begin{align}\label{eqn:gammaDDPCbias}
        b_{\gamma}(\gamma) := \begin{bmatrix}
            \beta L_{VV} & L_{Y \Epsilon}
        \end{bmatrix} \gamma,
    \end{align}
    and the DeePC Bias as a function of $g$ is 
    \begin{align}\label{eqn:DeePCbias}
        b_{g}(g) := \big(
            \beta_V + L_{Y \Epsilon} Q_{\Epsilon}
        \big) g.
    \end{align}
\end{lemma}


Lemma \ref{lemma:DDPCbias} makes it evident that $\gamma$-DDPC/DeePC bias can be decomposed into Subspace Bias, equivalent to the $\beta v$ bias described in \eqref{eqn:biasDef}, and \textit{Optimism Bias}.
\begin{definition}
    The (LTI) \emph{Optimism Bias} for a given $\gamma$ is $L_{Y \Epsilon} \gamma_{\epsilon}$, and for a given $g$ is $L_{Y \Epsilon} Q_\Epsilon g$.
\end{definition}
The Optimism Bias arises from the DDPC optimization ``optimistically believing'' that the innovation $\epsilon_f$ will help, rather than acting randomly as noise.
The Optimism Bias can also be interpreted as the prediction bias introduced by putting a ``slack'' (relaxation) on the output prediction \cite{berberich2020data}.

\begin{remark}\label{rem:ceError}
    For an LTI system with 
    open-loop training data satisfying Assumption \ref{ass:peristency}, $S = H$ 
    and the bias of the $\gamma$-DDPC prediction $y_f = \begin{bmatrix}L_{Y V} & L_{Y \Epsilon} \end{bmatrix} \gamma$ (the DeePC prediction $y_f = Y_f g$) is the Optimism Bias $L_{Y \Epsilon} \gamma_{\epsilon}$ ($L_{Y \Epsilon} Q_\Epsilon g$).
\end{remark}
\noindent Remark \ref{rem:ceError} makes it evident why $\gamma_{\epsilon}$ is often set to 0 in $\gamma$-DDPC and why a large Projection Regularizer \cite{dorfler2022bridging}, which has the same effect as setting $\gamma_{\epsilon} \!= \!0$, is often used for DeePC.

\begin{remark}
    SPC is a particular case of $\gamma$-DDPC for which $\gamma_{\epsilon} = 0$ and $\gamma_v$ is not regularized, and thus Lemma \ref{lemma:spcBiasLinComb} is a particular case of Lemma \ref{lemma:DDPCbias}. 
\end{remark}

In general, however, $\gamma_{\epsilon}$ is not set to 0 in $\gamma$-DDPC as it may sometimes 
be advantageous to trade off Optimism Bias for optimization cost \cite{dorfler2022bridging, breschi2023uncertainty}. 
In addition, the Optimism Bias is defined for DDPC applied to LTI systems, not nonlinear systems, for which $\gamma_\epsilon \neq 0$ may be advantageous \cite{dorfler2022bridging}.

Similar to Theorem \ref{thm:bias}, which applies to SPC, Theorem \ref{thm:biasDDPC} below quantifies the maximum bias of $\gamma$-DDPC and DeePC.

\begin{theorem}\label{thm:biasDDPC} 
    $\\$ 
    Consider an LTI system, i.e., \eqref{eqn:sysDef}, 
    
    \noindent a) If training data is collected in open loop,
    \begin{align*}
        \begin{aligned} 
        &\underset{\zeta}{\text{max}} \;\; \left\lVert b_\zeta(\zeta) \right\lVert \\
        &\text{s.t.} \;\ ||\zeta|| = 1  
        \end{aligned} 
    \quad = \sigma_\text{max} \left( L_{Y \Epsilon} \right) \neq 0,
    \end{align*}
    where $\zeta$ is $\gamma$ for $\gamma$-DDPC, or $g$ for DeePC.

    \noindent b) If the training data is gathered using feedback, 
    \begin{align*}
        \begin{aligned} 
        &\underset{\zeta}{\text{max}} \;\; \left\lVert b_\zeta(\zeta) \right\lVert \\
        &\text{s.t.} \;\ ||\zeta|| = 1  
        \end{aligned} 
    \quad = \sigma_\text{max} \left( \begin{bmatrix}
        \beta  L_{VV} & L_{Y \Epsilon}
    \end{bmatrix} \right) \neq 0,
    \end{align*}
    where $\zeta$ is $\gamma$ for $\gamma$-DDPC, or $g$ for DeePC.
\end{theorem}

Theorem \ref{thm:biasDDPC} bounds the maximum bias for $\gamma$-DDPC (DeePC) for a given $\gamma$ ($g$). With open-loop training data, the maximum bias depends only on the Optimism Bias matrix $L_{Y \Epsilon}$, whereas with closed-loop data the maximum bias also depends on the Subspace Bias Predictor $\beta$ defined in \eqref{eqn:betaExpression}.

\section{Consistent Multistep Prediction}\label{sec:TransientPredictor}

The Transient Predictor Method for estimating the Multistep Predictor \eqref{eqn:optMultiPred},  introduced in \cite{moffat2024transient}, provides $\widehat{H}$, a consistent estimate of $H$, regardless of whether the training data was gathered in open or closed-loop.
The Transient Predictor Method accomplishes this by first estimating $\Phi$, the bank of increasing-length, ``transient'' single-step predictors, and then uses $\widehat{\Phi}$ to estimate $\widehat{H}$.
The nonzero portion of each block-row of $\widehat{\Phi}$, corresponding to 
future timestep $k$, has the form
\begin{align}\label{eqn:singleStepPred}
    \widehat{\Phi}_{k} = Y_{[l+1,l+1]} Z_{[1,l]}^T \left(Z_{[1,l]} Z_{[1,l]}^T\right)^{-1} \in \mathbb{R}^{q\times(q+m)l},
\end{align}
where $k \in [1,\tau]$ and $l = k+\rho-1$.
Under Assumption \ref{ass:peristency}, \eqref{eqn:singleStepPred} is a consistent predictor of the true single-step predictor with $l$ lead-in data.
By estimating $\widehat{\Phi}$ as an intermediary step before estimating $\widehat{H}$, the Transient Predictor Method does not allow for the correlation between the innovation data $\Epsilon_f$ and the input data $U_f$ to affect the prediction of $y_f$. 

\subsection{Estimating the Subspace Bias Predictor $\beta$ from data}\label{sec:estimatingB}
$\widehat{\Phi} \! \in \! \mathbb{R}^{q\tau\times(q+m)(\rho+\tau)}$ can be split up into $\widehat{\Phi}_p \! \in \! \mathbb{R}^{q\tau \times (q+m)\rho}\!$, $\widehat{\Phi}_y \! \in \! \mathbb{R}^{q\tau \times q\tau}\!$, and $\widehat{\Phi}_u \! \in \! \mathbb{R}^{q\tau \times m\tau}\!$, as done in \cite{moffat2024transient}, such that
\begin{align*}
    \widehat{H} = \begin{bmatrix}
    \left(I - \widehat{\Phi}_{y}\right)^{-1}\widehat{\Phi}_p & \left(I - \widehat{\Phi}_{y}\right)^{-1}\widehat{\Phi}_{u}
    \end{bmatrix}.
\end{align*}
The Subspace Bias Predictor estimated from data is
\begin{align}\label{eqn:biasEstimate}
    \widehat{\beta} = \left(I - \widehat{\Phi}_y \right)^{-1}\widehat{\Epsilon}_f U_f^T \widehat{M} v \in \mathbb{R}^{(q \tau)},
\end{align}
where \eqref{eqn:Mdef} gives $\widehat{M}$ and the sample innovation estimate
$$
    \widehat{\Epsilon}_f = Y_f - \widehat{\Phi} Z_{[1,\rho+\tau]}.
$$



\section{Experiments}\label{sec:experiments}

To investigate the impact of prediction bias on DDPC we conducted experiments on the double integrator, the Euler Discretization of $y = \ddot{x}$, with Gaussian measurement noise $\sim \mathcal{N}(0,0.0001)$.
We compared SPC, Transient Predictive Control (TPC), which uses $\widehat{H}$ from \cite{moffat2024transient}/Section \ref{sec:TransientPredictor}, and ``2norm-DDPC,'' which runs the following optimization: 
\begin{align}
    \underset{u_f, y_f, \gamma}{\min} \quad & J(u_f,y_f) + \lambda \|\gamma\| \label{eqn:2normDDPC}\\
    \text{s.t.} \quad \quad &\begin{bmatrix}
                    v \\ y_f
                \end{bmatrix} =
                \begin{bmatrix}
                    L_{VV} & 0 \\
                    L_{Y V} & L_{Y \Epsilon}
                \end{bmatrix}
                \gamma. \nonumber
\end{align} 
2norm-DDPC is $\gamma$-DDPC with just a 2-norm $\gamma$ regularization, and is equivalent to DeePC with just a 2-norm $g$ regularization. 
We use ``2norm-DDPC'' as $\gamma$-DDPC and DeePC suggest additional regularizations/settings such as setting $\gamma_\epsilon = 0$ for $\gamma$-DDPC or using a large Projection Regularization for DeePC \cite{dorfler2022bridging}. 
As the focus of this paper is bias and not regularization/variance, we only include regularization for 2norm-DDPC, which is ill-posed with noisy data and no regularization, while noting that SPC and TPC also benefit from regularization.

The 10-step trajectory was determined by optimizing 
$J(u_f) = \left\Vert h(\cdot) - y_r \right\Vert^2_Q + \left\Vert u_f \right\Vert^2_R,$
where $y_r$ is the output reference which is 1 for position $y_1$ and 0 for velocity $y_2$, $R=I_\tau$, and $Q=I_\tau \otimes Q_1$, where $Q_1 = \diag(1000,10)$.
$h(\cdot)$ is $\widehat{S}v$ for SPC, $\begin{bmatrix} L_{Y V} & L_{Y \Epsilon} \end{bmatrix} \gamma$ for 2norm-DDPC, and 
$\widehat{H} v$ for TPC.
$\rho = 10$ lead-in measurements were used, all equal to 0 as the $t=0$ initial position of the mass was 0. 

The training data set consisted of 1000 training data points collected either in closed or open loop.
The open-loop data was gathered by perturbing the system with random Gaussian excitation $\sim \mathcal{N}(0,0.01)$.
The closed-loop data was gathered by perturbing the system with the same proportional-derivative feedback as in Section \ref{sec:SPCbias} ($f\big(y(t)\big) = -2.5 y_1(t) - 3y_2(t)$) and additive Gaussian excitation $\sim \mathcal{N}(0,0.01)$, which resulted in a Noise-to-Signal Ratio (NSR) of $0.5\%$. 


\subsection{Open-Loop Double-Integrator Experiment}\label{sec:OLexperiments}

\begin{figure}
\begin{subfigure}{0.48\columnwidth}
\includegraphics[width=\textwidth]{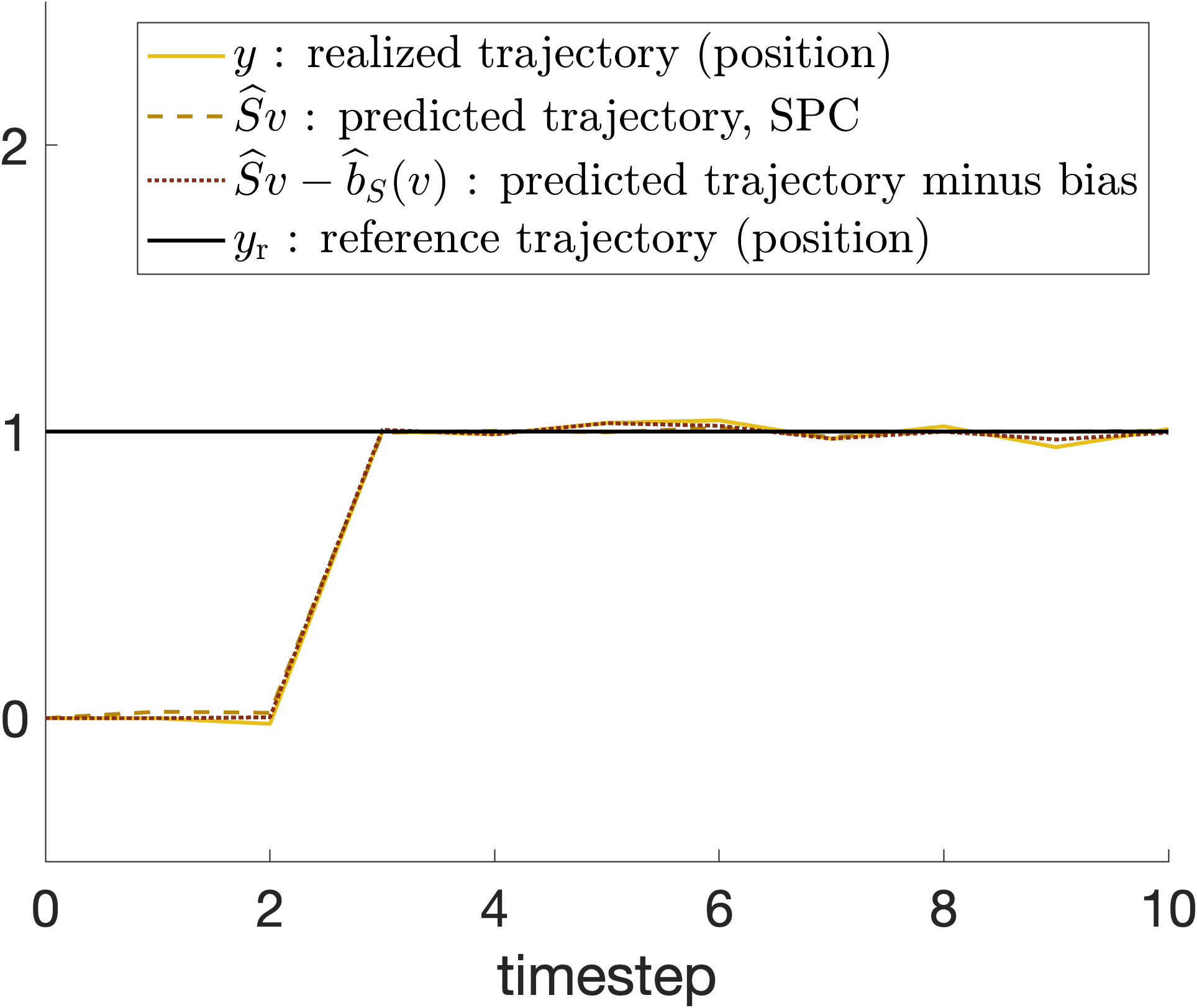}
\caption{\footnotesize SPC, open-loop data\\ \smaller(No regularization used)\normalsize} 
\label{fig:SPC_openloop_OLdata}
\end{subfigure}\hspace*{\fill}
\begin{subfigure}{0.48\columnwidth}
\includegraphics[width=\textwidth]{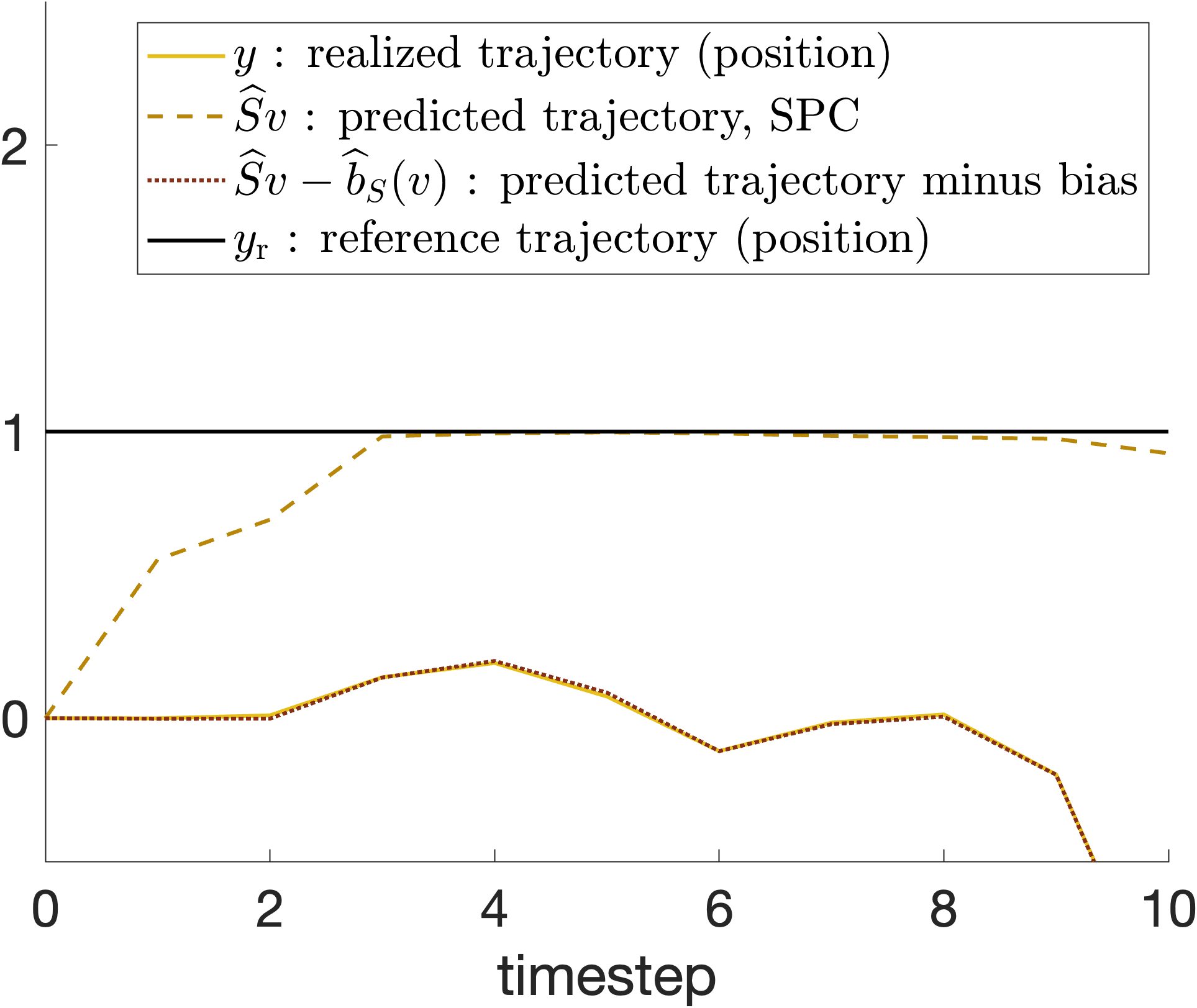}
\caption{\footnotesize SPC, closed-loop data\\ \smaller(No regularization used)\normalsize} 
\label{fig:SPC_openloop_CLdata}
\end{subfigure}

\medskip
\begin{subfigure}{0.48\columnwidth}
\includegraphics[width=\textwidth]{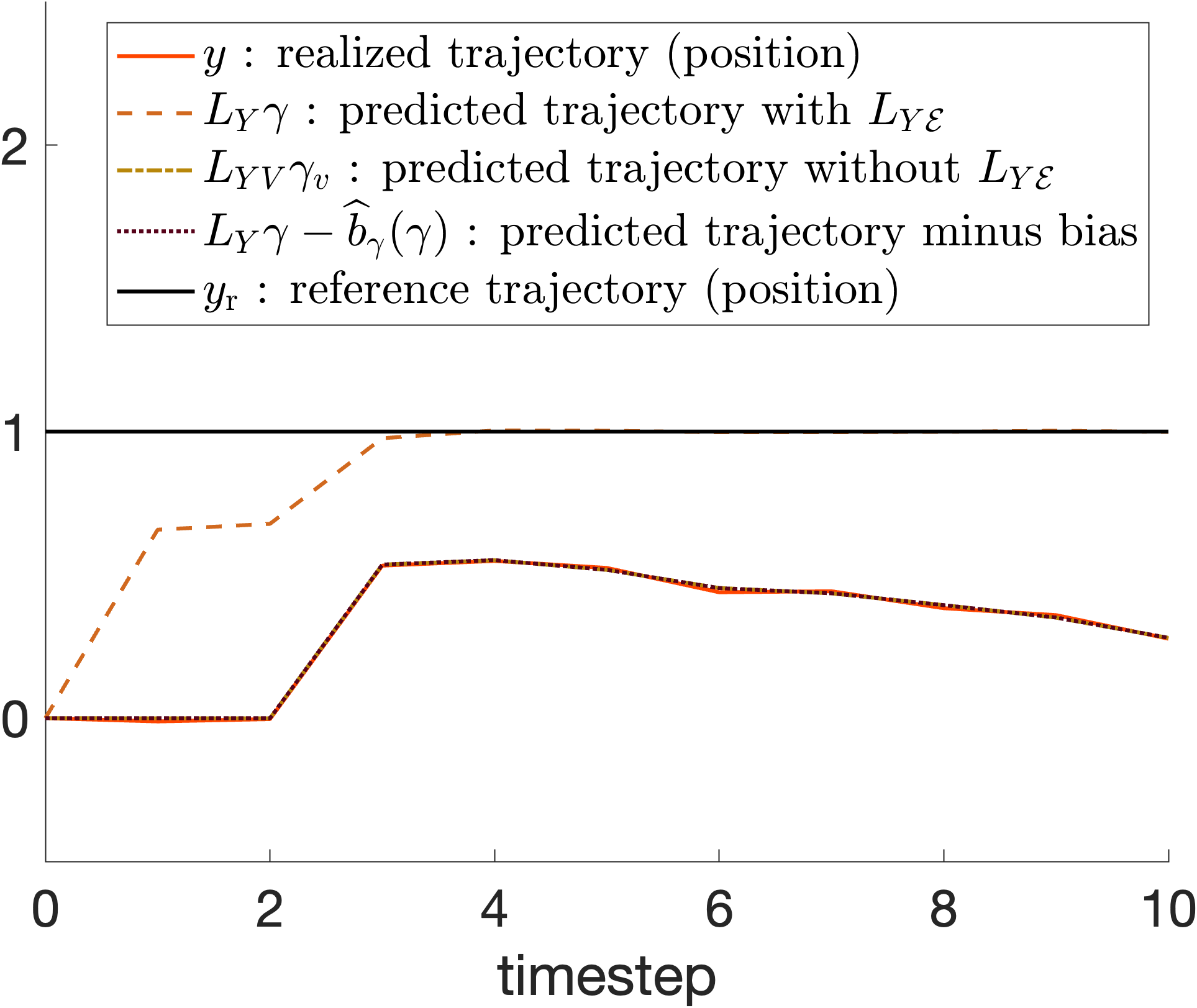}
\caption{\footnotesize 2norm-DDPC, 
open-loop data\\ \smaller($0.1||\gamma||^2$ regularization used)\normalsize} 
\label{fig:fullgamma_openloop_OLdata}
\end{subfigure}\hspace*{\fill}
\begin{subfigure}{0.48\columnwidth}
\includegraphics[width=\textwidth]{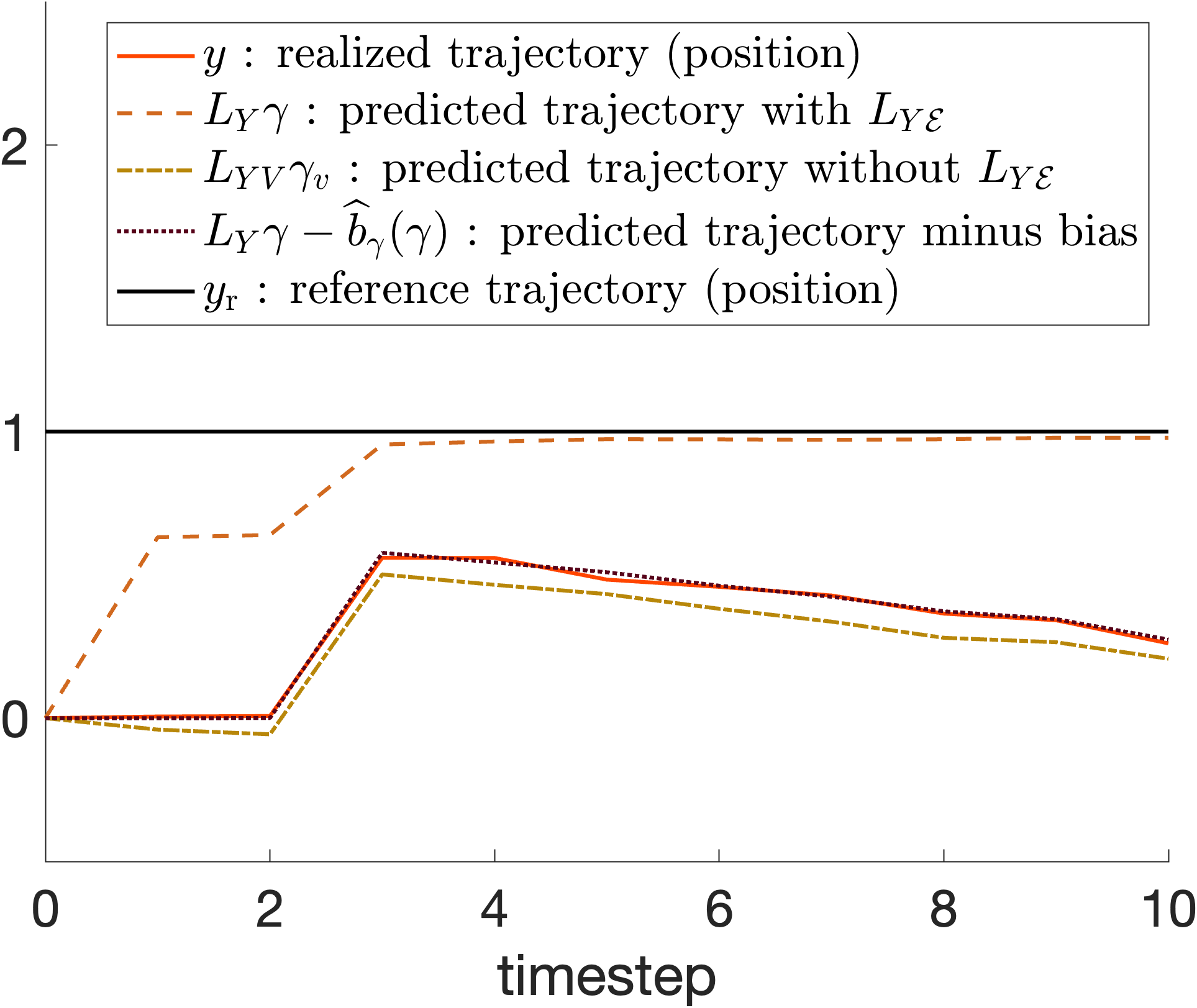}
\caption{\footnotesize 2norm-DDPC, 
closed-loop data\\ \smaller($0.1||\gamma||^2$ regularization used)\normalsize} 
\label{fig:fullgamma_openloop_CLdata}
\end{subfigure}

\medskip
\begin{subfigure}{0.48\columnwidth}
\includegraphics[width=\textwidth]{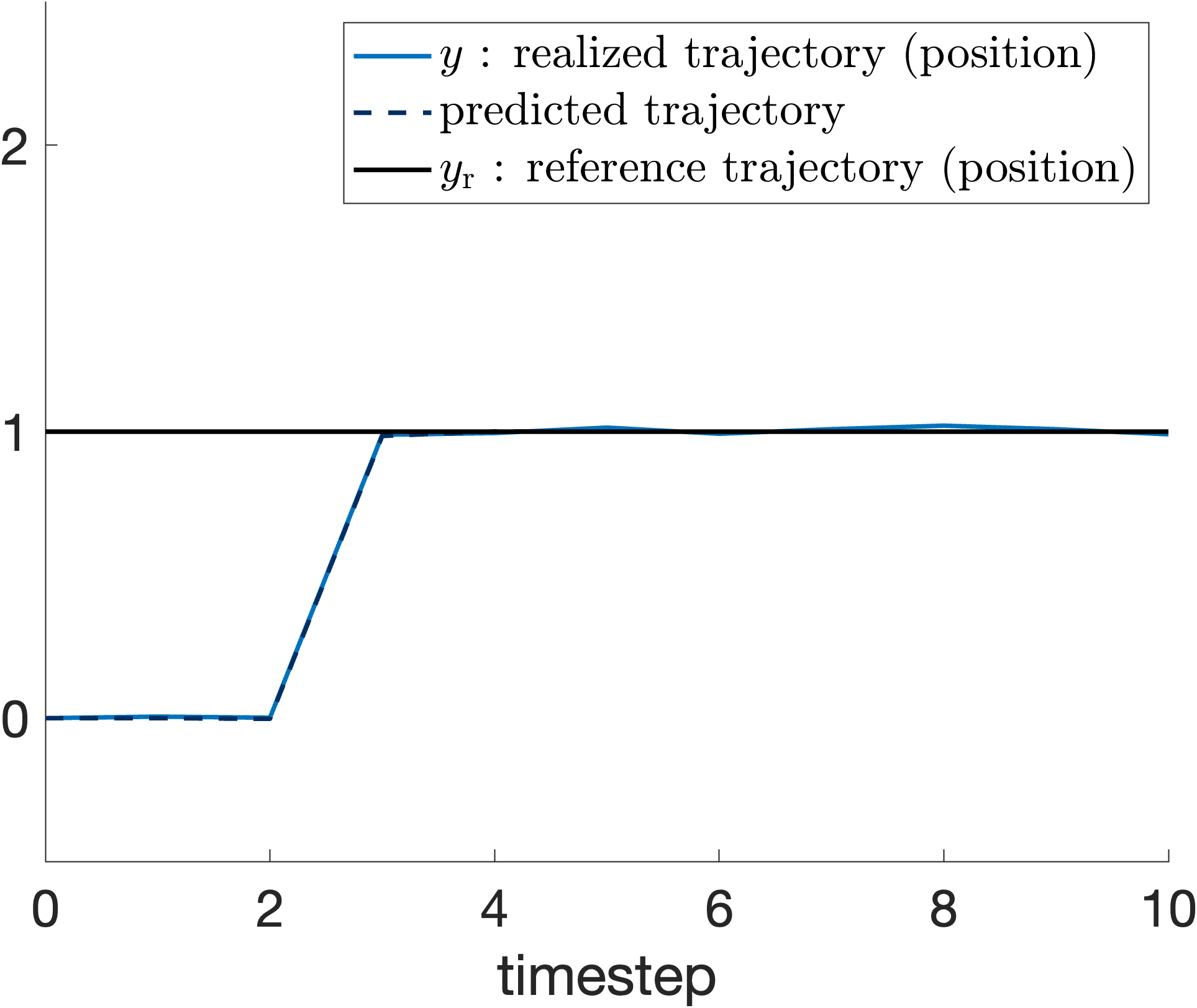}
\caption{\footnotesize TPC, open-loop data\\ \smaller(No regularization used)\normalsize} 
\label{fig:TPC_openloop_OLdata}
\end{subfigure}\hspace*{\fill}
\begin{subfigure}{0.48\columnwidth}
\includegraphics[width=\textwidth]{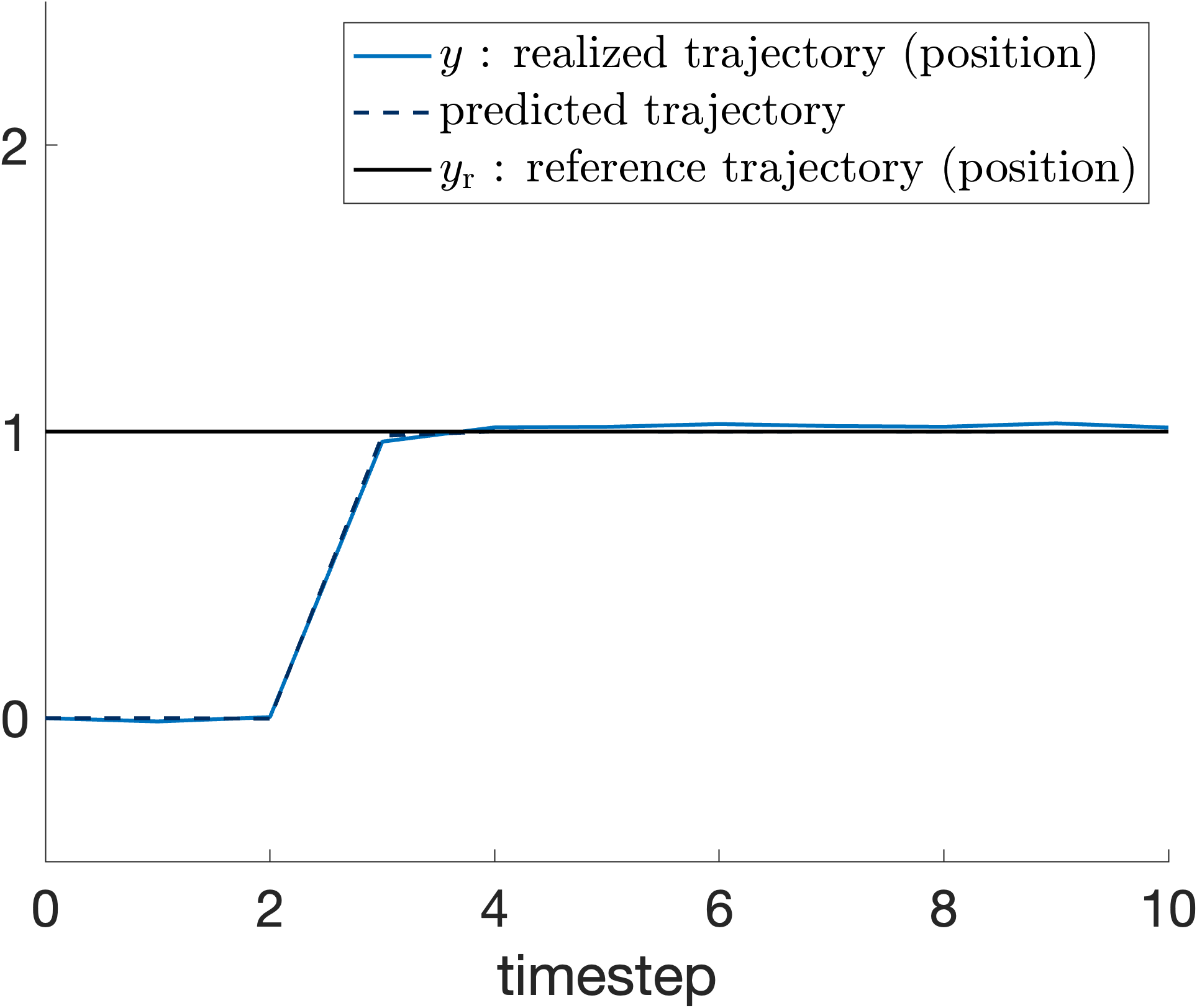}
\caption{\footnotesize TPC, closed-loop data\\ \smaller(No regularization used)\normalsize}
\label{fig:TPC_openloop_CLdata}
\end{subfigure}

\caption{\small Planned trajectories vs. open-loop realized trajectories} 
\label{fig:OLtrajectories}
\end{figure}

The plots for six open-loop tests, in which the 10-step input sequence produced by the DDPC optimization is applied to the system without re-optimizing, are in Fig. \ref{fig:OLtrajectories}.
Three insights can be drawn from the plots:
\begin{enumerate}
    \item SPC performs well with open loop data but not with closed-loop data. This drop in performance is due to the Subspace Bias. 
    \item 2norm-DDPC is primarily affected by ($L_{Y \Epsilon}$) Optimism Bias with both open and closed-loop data. With closed-loop data, there is an additional bias which is explained by DDPC's Subspace Bias. 
    \item TPC performs well with both open and closed-loop data. 
\end{enumerate}

\subsection{Closed-Loop Double-Integrator Experiment}

\begin{figure}
\begin{subfigure}{0.48\columnwidth}
\includegraphics[width=\textwidth]{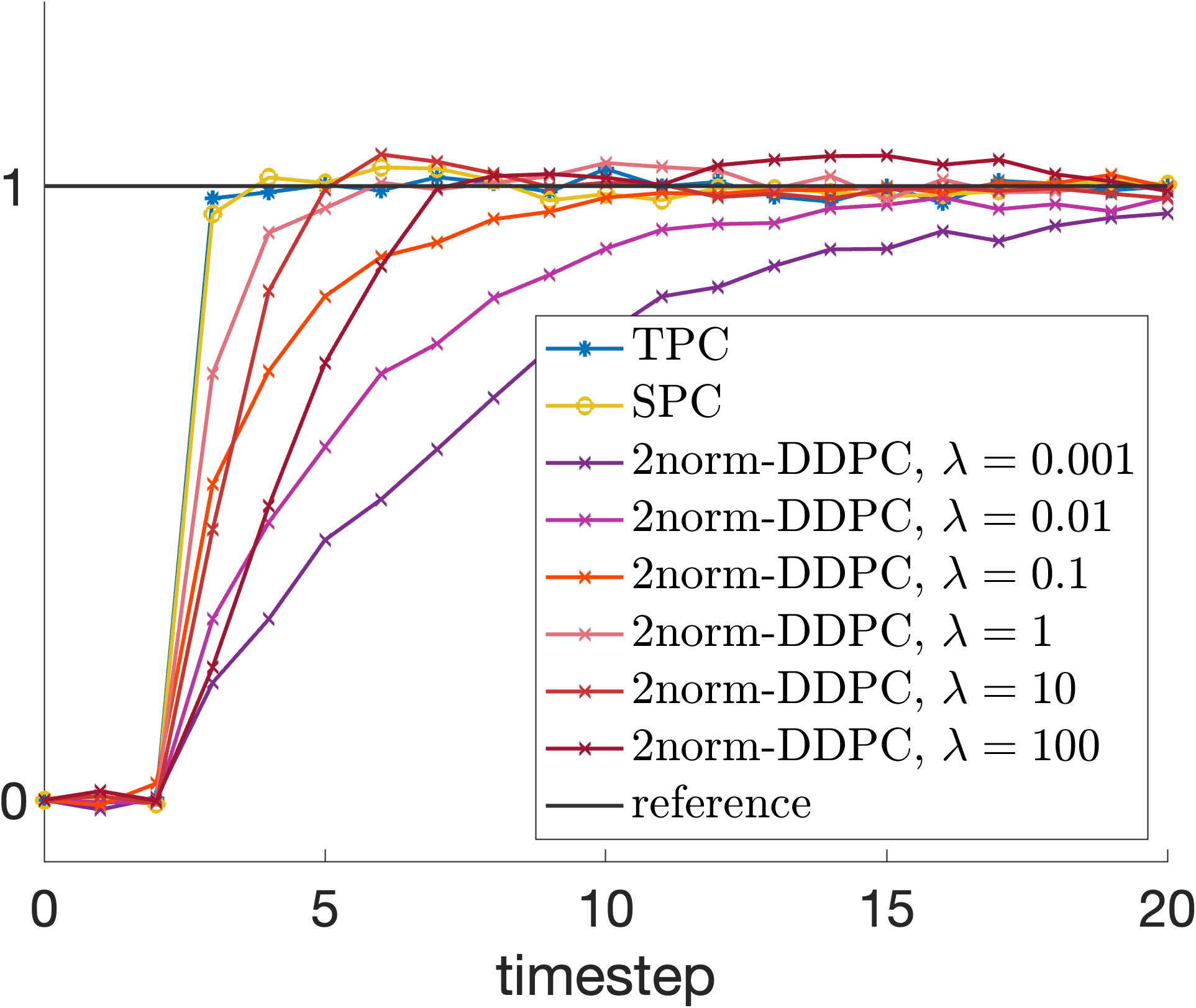}
\caption{Open-loop data} 
\label{fig:closedloop_OLdata}
\end{subfigure}\hspace*{\fill}
\begin{subfigure}{0.48\columnwidth}
\includegraphics[width=\textwidth]{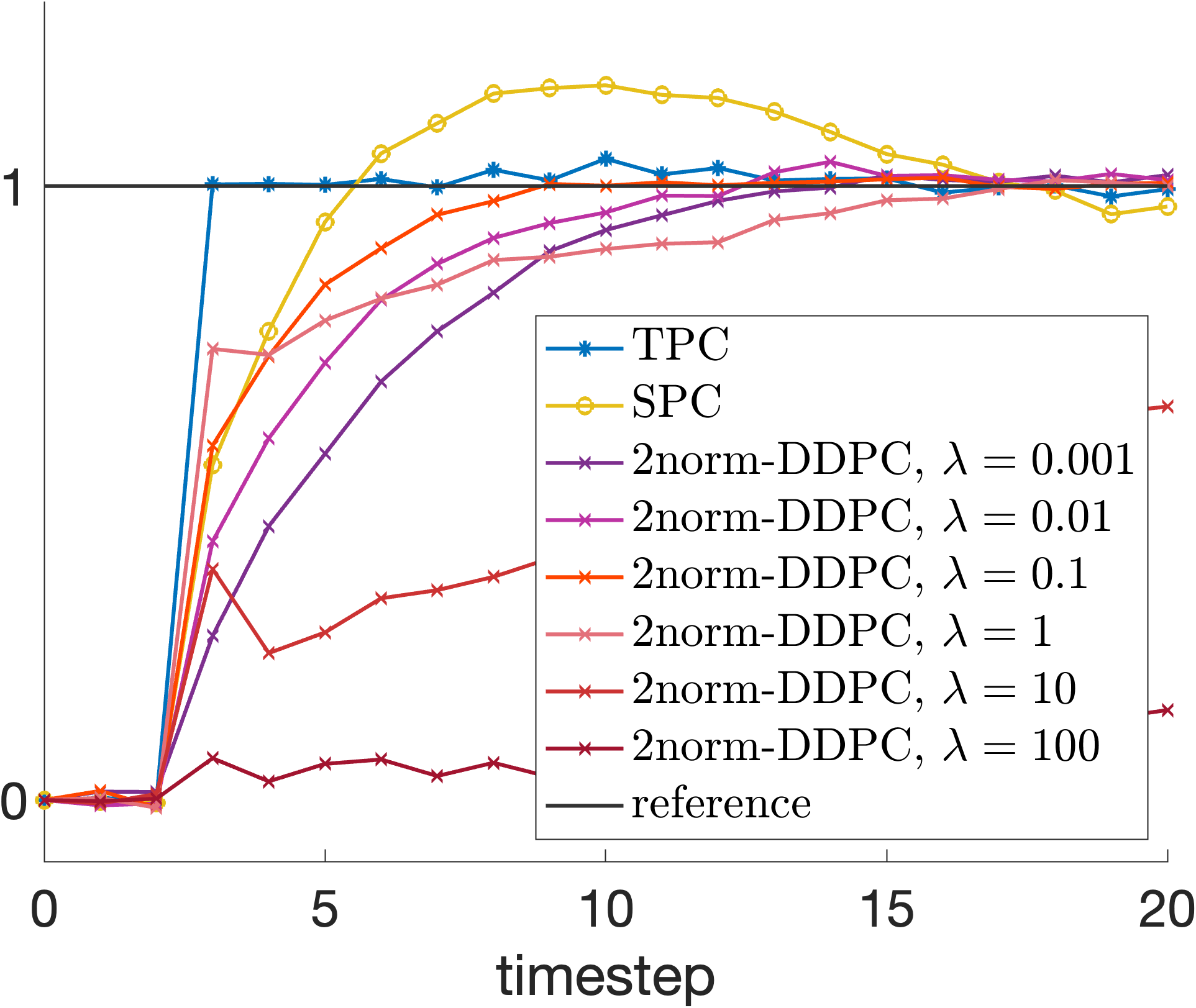}
\caption{Closed-loop data} 
\label{fig:closedloop_CLdata}
\end{subfigure}

\caption{\small Realized closed-loop trajectories (position)} \label{fig:CLtrajectories}
\end{figure}

Fig. \ref{fig:CLtrajectories} includes the same six tests as Fig. \ref{fig:OLtrajectories}, but with the DDPC optimization run in closed-loop with the system, i.e., replanning at each timestep.
For 2norm-DDPC, 
Fig. \ref{fig:CLtrajectories} plots the performance for six different $\lambda$ values logarithmically swept from 0.001 to 100.
Fig. \ref{fig:closedloop_OLdata}, which compares the performances with open-loop training data, and Fig. \ref{fig:closedloop_CLdata}, which compares the performances with closed-loop training data, corroborate insights (1)-(3) from Section \ref{sec:OLexperiments}. 

It is evident that 2norm-DDPC \eqref{eqn:2normDDPC} is slower to reach the reference than TPC regardless of the regularization value. 
For large $\lambda$, 2norm-DDPC is slow to reach the reference due to the $\gamma$ regularization indirectly penalizing $u_f$ in addition to $\left\Vert u_f \right\Vert^2_R$. 
For small $\lambda$, 2norm-DDPC is slow to reach the reference due to Optimism Bias, which leads 2norm-DDPC to believe that it will converge with less control effort than is actually needed.


\section{Conclusion}

This paper quantified the Subspace Bias and the Optimism Bias of subspace-based DDPC methods and demonstrated that they impact control performance.
It also demonstrated that Transient Predictive Control, which does not suffer from Subspace Bias or Optimism Bias and is consistent regardless of how the training data are gathered, performs well with both open and closed-loop data.

\bibliographystyle{IEEEtran}
\bibliography{biblio}

\begin{thebibliography}{10}
\providecommand{\url}[1]{#1}
\csname url@samestyle\endcsname
\providecommand{\newblock}{\relax}
\providecommand{\bibinfo}[2]{#2}
\providecommand{\BIBentrySTDinterwordspacing}{\spaceskip=0pt\relax}
\providecommand{\BIBentryALTinterwordstretchfactor}{4}
\providecommand{\BIBentryALTinterwordspacing}{\spaceskip=\fontdimen2\font plus
\BIBentryALTinterwordstretchfactor\fontdimen3\font minus \fontdimen4\font\relax}
\providecommand{\BIBforeignlanguage}[2]{{%
\expandafter\ifx\csname l@#1\endcsname\relax
\typeout{** WARNING: IEEEtran.bst: No hyphenation pattern has been}%
\typeout{** loaded for the language `#1'. Using the pattern for}%
\typeout{** the default language instead.}%
\else
\language=\csname l@#1\endcsname
\fi
#2}}
\providecommand{\BIBdecl}{\relax}
\BIBdecl

\bibitem{coulson2019data}
J.~Coulson, J.~Lygeros, and F.~D{\"o}rfler, ``Data-enabled predictive control: In the shallows of the {DeePC},'' in \emph{2019 18th European Control Conference (ECC)}.\hskip 1em plus 0.5em minus 0.4em\relax IEEE, 2019, pp. 307--312.

\bibitem{berberich2020data}
J.~Berberich, J.~K{\"o}hler, M.~A. M{\"u}ller, and F.~Allg{\"o}wer, ``Data-driven model predictive control with stability and robustness guarantees,'' \emph{IEEE Transactions on Automatic Control}, vol.~66, no.~4, pp. 1702--1717, 2020.

\bibitem{coulson2021distributionally}
J.~Coulson, J.~Lygeros, and F.~D{\"o}rfler, ``Distributionally robust chance constrained data-enabled predictive control,'' \emph{IEEE Transactions on Automatic Control}, vol.~67, no.~7, pp. 3289--3304, 2021.

\bibitem{huang2023robust}
L.~Huang, J.~Zhen, J.~Lygeros, and F.~D{\"o}rfler, ``Robust data-enabled predictive control: Tractable formulations and performance guarantees,'' \emph{IEEE Transactions on Automatic Control}, vol.~68, no.~5, pp. 3163--3170, 2023.

\bibitem{breschi2022unified}
V.~Breschi, A.~Chiuso, and S.~Formentin, ``Data-driven predictive control in a stochastic setting: a unified framework,'' \emph{Automatica}, vol. 152, no. 110961, 2023.

\bibitem{chiuso2025harnessing}
A.~Chiuso, M.~Fabris, V.~Breschi, and S.~Formentin, ``Harnessing uncertainty for a separation principle in direct data-driven predictive control,'' \emph{Automatica}, vol. 173, p. 112070, 2025.

\bibitem{FAVOREEL1999}
W.~Favoreel, B.~D. Moor, and M.~Gevers, ``{SPC}: Subspace predictive control,'' \emph{IFAC Proceedings Volumes}, vol.~32, no.~2, pp. 4004--4009, 1999, 14th IFAC World Congress 1999, Beijing, Chia, 5-9 July.

\bibitem{dong2008CLSPC}
J.~Dong, M.~Verhaegen, and E.~Holweg, ``Closed-loop subspace predictive control for fault tolerant mpc design,'' \emph{IFAC Proceedings Volumes}, vol.~41, no.~2, pp. 3216--3221, 2008.

\bibitem{huang2008dynamic}
B.~Huang and R.~Kadali, \emph{Dynamic modeling, predictive control and performance monitoring: a data-driven subspace approach}.\hskip 1em plus 0.5em minus 0.4em\relax Springer, 2008.

\bibitem{dinkla2023bias}
R.~Dinkla, S.~P. Mulders, J.-W. van Wingerden, and T.~Oomen, ``Closed-loop aspects of data-enabled predictive control,'' \emph{IFAC-PapersOnLine}, vol.~56, no.~2, pp. 1388--1393, 2023.

\bibitem{Vanov-book}
P.~{Van~Overschee} and B.~{De~Moor}, \emph{Subspace Identification for Linear Systems}.\hskip 1em plus 0.5em minus 0.4em\relax Kluwer Academic Publications, 1996.

\bibitem{ChiusoAUTO06}
A.~Chiuso, ``The role of {V}ector {A}uto{R}egressive modeling in predictor based subspace~identification,'' \emph{Automatica}, vol.~43, no.~6, pp. 1034--1048, June 2007.

\bibitem{vanderveen2013closed}
G.~Van~der Veen, J.-W. van Wingerden, M.~Bergamasco, M.~Lovera, and M.~Verhaegen, ``Closed-loop subspace identification methods: an overview,'' \emph{IET Control Theory \& Applications}, vol.~7, no.~10, pp. 1339--1358, 2013.

\bibitem{LjungMcK1996}
L.~Ljung and T.~McKelvey, ``Subspace identification from closed loop data,'' \emph{Signal Processing}, vol.~52, no.~2, pp. 209--216, 1996.

\bibitem{OverscheeDeMoor97}
P.~{Van~Overschee} and B.~{De~Moor}, ``Closed loop subspace systems identification,'' in \emph{Proc. of 36th IEEE Conference on Decision \& Control}, San Diego, CA, 1997, pp. pp. 1848--1853.

\bibitem{ChiusoP-05a}
A.~Chiuso and G.~Picci, ``Consistency analysis of some closed-loop subspace identification methods,'' \emph{Automatica}, vol.~41, no.~3, pp. 377--391, 2005.

\bibitem{pouliquen2010indirect}
M.~Pouliquen, O.~Gehan, and E.~Pigeon, ``An indirect closed loop subspace identification method,'' in \emph{49th IEEE Conference on Decision and Control (CDC)}.\hskip 1em plus 0.5em minus 0.4em\relax IEEE, 2010, pp. 4417--4422.

\bibitem{QinL2003a}
S.~Qin and L.~Ljung, ``Closed-loop subspace identification with innovation estimation,'' in \emph{Proc. of SYSID 2003}, Rotterdam, 2003.

\bibitem{moffat2024transient}
K.~Moffat, D.~Florian, and A.~Chiuso, ``{The Transient Predictor},'' in \emph{2024 IEEE 63rd Conference on Decision and Control (CDC)}.\hskip 1em plus 0.5em minus 0.4em\relax IEEE, 2024.

\bibitem{dinkla2024closed}
R.~Dinkla, S.~Mulders, T.~Oomen, and J.-W. van Wingerden, ``Closed-loop data-enabled predictive control and its equivalence with closed-loop subspace predictive control,'' \emph{arXiv preprint arXiv:2402.14374}, 2024.

\bibitem{dorfler2022bridging}
F.~Dorfler, J.~Coulson, and I.~Markovsky, ``Bridging direct \& indirect data-driven control formulations via regularizations and relaxations,'' \emph{IEEE Transactions on Automatic Control}, 2022.

\bibitem{breschi2023uncertainty}
V.~Breschi, M.~Fabris, S.~Formentin, and A.~Chiuso, ``Uncertainty-aware data-driven predictive control in a stochastic setting,'' \emph{IFAC-PapersOnLine}, vol.~56, no.~2, pp. 10\,083--10\,088, 2023.

\end{thebibliography}

\appendices



\section{Proofs}\label{app:Proofs}

\subsection*{Proof of Lemma \ref{lemma:SPCBias}}
\begin{proof}[\unskip\nopunct]
    \begin{align*}
        \beta &= \mathbb{E}\left[Y_f V^T\right] \mathbb{E}\left[V V^T\right]^{-1}  - H \\
        &= \mathbb{E}\left[\left(H V  + H_\epsilon \Epsilon_f\right) V^T\right] \mathbb{E}\left[V V^T\right]^{-1} - H  \\
        &= H_\epsilon \mathbb{E}\left[\Epsilon_f V^T\right] \mathbb{E}\left[V V^T\right]^{-1}
    \end{align*}
    Defining
    \begin{align}
            \Omega &:= \mathbb{E}\left[U_f U_f^T \right] - \mathbb{E}\left[U_f Z_p^T \right] \mathbb{E}\left[ Z_p Z_p^T \right]^{-1} \mathbb{E}\left[Z_p U_f^T \right] \text{, and} \nonumber \\
            M &:= \begin{bmatrix}
            - \Omega^{-1} \mathbb{E}\left[U_f Z_p^T \right] \mathbb{E}\left[ Z_p Z_p^T \right]^{-1} & \Omega^{-1} \label{eqn:Mdef}
            \end{bmatrix}, 
    \end{align}
    and using the fact that $\mathbb{E}\left[\Epsilon_f Z_p^T\right] = 0$, we have
    \begin{align*}
        \beta &= H_\epsilon \begin{bmatrix}
            0 & \mathbb{E}\big[\Epsilon_f U_f^T\big]
        \end{bmatrix} \begin{bmatrix}
            \mathbb{E}\left[Z_p Z_p^T \right] & \mathbb{E}\left[Z_p U_f^T \right] \\
            \mathbb{E}\left[U_f Z_p^T \right] & \mathbb{E}\left[U_f U_f^T \right] 
        \end{bmatrix}^{-1} \\
        &= H_\epsilon \mathbb{E}\big[\Epsilon_f U_f^T\big] M
    \end{align*}
    from the formula for the inverse of $2 \times 2$ block matrix.
\end{proof}

\subsection*{Proof of Lemma \ref{lemma:spcBiasLinComb}}
\begin{proof}[\unskip\nopunct] 
    $b_S(v) = \beta v = \beta V g = \beta_V g.$
\end{proof}

\subsection*{Proof of Theorem \ref{thm:bias}}
\begin{proof}[\unskip\nopunct]
    (a) follows from Lemma \ref{lemma:SPCBias}, and that $\mathbb{E}\big[\Epsilon_f U_f^T\big] = 0$ when the training data is gathered in open loop.
    
    The first statement in (b) follows from Lemma \ref{lemma:spcBiasLinComb} and the definition of $\sigma_\text{max}$.
    
    The second statement in (b) follows from 
    \begin{align*}
        b_{S}(v) &= H_\epsilon \begin{bmatrix}
            0 & \mathbb{E}\big[\Epsilon_f U_f^T\big]
        \end{bmatrix} \mathbb{E}\left[ V V^T \right]^{-1/2} \mathbb{E}\left[ V V^T \right]^{-T/2} v \\
        &= A \xi \text{, where } \xi := \mathbb{E}\left[ V V^T \right]^{-T/2} v \text{, and } \mathbb{E}\left[ \xi \xi^T \right] = I
    \end{align*}
    if $v$ is drawn from the same distribution as $V$.
    \begin{align*}
        \mathbb{E}\left[ \left\Vert b_{S}(v)\right\Vert^2 \right] &= \mathbb{E}\left[\left(\xi^T A^T A \xi \right) \right]\\ &=\mathbb{E}\left[\Tr\left(A^T A \xi \xi^T \right) \right]\\
        &= \Tr\left( A^T A \right) \\ 
        &= \left\Vert H_\epsilon \begin{bmatrix}
            0 & \mathbb{E}\big[\Epsilon_f U_f^T\big]
        \end{bmatrix} \mathbb{E}\left[ V V^T \right]^{-1/2}\right\Vert^2_F 
    \end{align*}
\end{proof}

\subsection*{Proof of Lemma \ref{lemma:DDPCbias}}

\begin{proof}[\unskip\nopunct]
    The Lemma follows directly from $\beta = S - H$ and the definitions of $b_{\gamma}(\gamma)$ and $b_{g}(g)$. For \eqref{eqn:gammaDDPCbias} we use $S = L_{YV} L_{VV}^{-1}$, and for \eqref{eqn:DeePCbias} we use $V = L_{VV} Q_V$ and $Y_f = L_{YV} Q_V + L_{Y\Epsilon} Q_\Epsilon$. 
\end{proof}

\subsection*{Proof of Theorem \ref{thm:biasDDPC}}

\begin{proof}[\unskip\nopunct]
    The Theorem follows from Lemma \ref{lemma:DDPCbias} and the definition of $\sigma_\text{max}$. For DeePC, $b_{g}(g) = \begin{bmatrix}
        \beta L_{VV} & L_{Y \Epsilon}
    \end{bmatrix} Q g$. As $Q$ is unitary, $\sigma_\text{max} \! \left( \begin{bmatrix}
        \beta L_{VV} & L_{Y \Epsilon}
    \end{bmatrix} Q \right) = \sigma_\text{max} \! \left( \begin{bmatrix}
        \beta L_{VV} & L_{Y \Epsilon}
    \end{bmatrix} \right)$.
    For (a), as with Theorem \ref{thm:bias}, $\beta = 0$ when the data is collected in open loop.
\end{proof}

\end{document}